\documentclass[twocolumn,prb,reprint,amsmath,amssymb,aps,showpacs]{revtex4-1}

\usepackage{graphicx}
\usepackage{dcolumn}
\usepackage{bm}
\usepackage{graphics}
\usepackage{epsfig}
\usepackage{epstopdf}
\begin{document}

\title{The plasma oscillations in a two-dimensional electron-hole liquid under influence Rashba spin-orbit coupling}
\author{E.V. Dumanov$^{1,2}$}
\affiliation{$^{1}$Institute of Applied Physics of the Academy of Sciences of Moldova, Academic Str. 5, Chisinau, MD2028, Republic of Moldova\\
$^{2}$Technical University of Moldova, bd. Stefan cel Mare 168, MD2004, Chisinau, Republic of Moldova}

\date{\today}

\begin{abstract}
The plasma oscillations of the two-dimensional electron-hole system in the presence of a Rashba spin-orbit coupling are studied. Only the intra-Landau level excitations are taken into account when the electrons and holes are situated on their lowest Landau levels, the filling factor $v^{2}$ being less than 1. The ground state of the two-dimensional electron-hole system is supposed to be the electron-hole liquid. The dispersion relations for the optical and acoustical plasmon modes were obtained. The acoustical plasmon branch has a linear dispersion law in the range of small wave vectors and monotonically increases with saturation at higher values of wave vectors. The optical plasmon branch has a quadratic dependence in the range of long wavelength and a monotonic increasing with saturation as in the case of acoustical branch.
\end{abstract}

\maketitle
\tableofcontents

\section{Introduction}
In the last decade, many theoretical works devoted to the study of plasma oscillations in three-(3D) and two-dimensional(2D) electron-hole structures. The plasma oscillations of the one component electron gas in the 3D bulk crystals [1] as well as in 2D layers characterized by the square of the frequencies $\omega _{p}^{2}(q)=\frac{4\pi {{e}^{2}}{{n}_{e}}}{{{\varepsilon }_{0}}m}$ and $\omega _{p}^{2}(q)=\frac{2\pi {{e}^{2}}{{n}_{S}}q}{{{\varepsilon }_{0}}m}$, where $n_{e}$ and ${{n}_{S}}$ are the corresponding electron densities. Das Sarma and Madhukar [2] considered the two-component 2D electron gas(2DEG). Two operators of the density fluctuations ${{\hat{\rho }}_{1}}(q)$ and ${{\hat{\rho }}_{2}}(q)$ corresponding to each layer combine in phase and in opposite phases forming the optical and acoustical plasmon oscillations with the frequencies $\omega_{OP}(q)\sim \sqrt{q}$ and $\omega_{AP}(q)\sim {q}$ in the range of small wave vectors q. Kasyan and coautors [3, 4] investigated the interconnected plasmon-phonon excitations in 2D structures. Excitations of the 2DEG in a strong perpendicular magnetic field are completely different to the case discussed above in that the kinetic energy is suppressed by the Landau quantization and magnetic field.

Girvin, MacDonald and Platzman [5] proposed a magnetoroton theory of collective elementary excitations in 2DEG in conditions of the fractional quantum Hall effect (FQHE). It occurs in low-disordered, high-mobility samples with partially filled lowest Landau levels with filling factor of the form $v=1/q$, where q is an integer $(q\ne 1)$.  In this case the excitation is a collective effect arising from the many-body correlations due to Coulomb interaction. Decisive progress has been achieved by Laughlin [6] toward understanding the nature of the many body ground state due his variational wave function. The theory of the collective excitation spectrum proposed by [5] is closed analogous to Feynman's theory of superfluid Helium [7]. The main Feynman's arguments lead to the conclusion that on general ground the low lying excitations of any system will include density waves. In the case of filled Landau level $v=1$ because of Pauli exclusion principle the lowest excitations are necessarily the cyclotron modes in which particles are excited in the next Landau levels. Just this case was studied by Kallin and Halperin [8]. But in the case of FQHE the lowest Landau level(LLL) is fractionally filled. The Pauli principle no longer excludes low-energy intra-Landau-level excitations. For the FQHE case the primary importance have the low-lying excitations, rather than the high-energy inter-Landau-level cyclotron modes [5]. The energy spectrum obtained in [5] has an energy gap at zero wave vector $k=0$. It exhibits a deep magnetoroton minimum at dimensionless wave vector $kl=1$, where $l$ is the magnetic length, quite analogous to the roton minimum in helium. Fertig [9] investigated the excitation spectrum of two-layer and three-layer electron systems. The case of two-layer system in a strong perpendicular magnetic field with filling factor $v=1/2$ of the LLLs of each layer was considered. The spontaneous coherence of two-component 2DEG was introduced, which is equivalent to the BCS-type ground state of the superconductor. It represents the coherent pairing of the conduction electrons in one layer with the holes in the same conduction band of another layer. Such unusual excitons are named as FQHE excitons. Their excitations are expected to have the energies $\hbar \omega (k)={{E}_{ex}}(k)-{{E}_{ex}}(0)$, where ${{E}_{ex}}(k)$ is the energy of exciton with the wave vector $\vec{k}$. Such result was obtained by Paquet, Rice and Ueda [10], who studied the electron-hole system in a single layer. In the asymmetric case, when the distance between the layers is different from zero, the energy spectrum of elementary excitations is characterized by linear dependence on the wave vector in the region of long wavelengths as well by the roton type behavior at the intermediary values of wave vectors [9]. In the symmetric case, when the distance between the layers vanishes and the Coulomb interactions inter-layers and intra-layers are the same, the linear region of the energy spectrum is transformed into the quadratic dependence. Such type energy spectrum of elementary excitations in two-layer electron system was discussed also in the Ref. [11].

Moskalenko and coauthors [12] investigated the intra-Landau level excitations of the 2D electron-hole liquid (EHL) under the influence strong perpendicular magnetic field. It was shown that such excitations are characterized by two branches of the energy spectrum. One of them is the acoustical plasmon type branch with the linear dispersion law in the range of small wave vectors and monotonically increasing with saturation behavior at higher wave vectors. The second branch of the elementary excitations is an optical-plasmon branch with quadratic dispersion law at small wave vectors with a roton-type dispersion at intermediary wave vectors and with a similar behavior as the acoustical branch at higher wave vectors.
In [13] was considered the plasmon oscillations in the 2DEG under influence the spin-orbit splitting.  It was shown that transitions between different spin states produce narrow absorption band in the degenerate 2DEG. As a result, the plasmon spectrum was modified in the new type of oscillations namely, a spin-plasmon polariton.
Xu and coautors [14] investigated influence the Rashba spin-orbit interaction (RSOI) on the fast-electron optical spectrum of the 2DEG. They found that for a spin-split 2DEG, the spectrum of optical absorption is mainly induced by plasmon excitation via inter-SO electronic transition.
In [15] was proposed a concept of surface plasmon-polariton amplification in the structure comprising interface between dielectric, metal, and asymmetric quantum well.

Our paper is organized as follows. In the section II the starting Hamiltonian is introduced and the motion equations for the operators and Green's function are deduced. In section III the energy spectrum of the collective elementary excitations is obtained. Section IV contains the conclusions.
\section{The Hamiltonian and equation of motion for the operators and Green's functions}
The Hamiltonian of the Coulomb interaction of the electrons and holes in the frame of the lowest Landau levels has the form [16]
\begin{eqnarray}
H=\frac{1}{2}\sum\limits_{\vec{Q}}W_{\vec{Q}}(\hat{\rho }_{e}^{R}(\vec{Q})\hat{\rho }_{e}^{R}(-\vec{Q})-\hat{N}_{e})+ \nonumber \\
+\frac{1}{2}\sum\limits_{\vec{Q}}W_{\vec{Q}}(\hat{\rho }_{h}^{R}(\vec{Q})\hat{\rho }_{h}^{R}(-\vec{Q})-\hat{N}_{h})-   \\
-\sum\limits_{\vec{Q}}W_{\vec{Q}}\hat{\rho }_{e}^{R}(\vec{Q})\hat{\rho }_{h}^{R}(-\vec{Q}) \nonumber
\end{eqnarray}
where
\begin{eqnarray}
W_{\vec{Q}}=V_{\vec{Q}}e^{-\frac{Q^{2}l^{2}}{2}}, \\
V_{\vec{Q}}=\frac{2\pi e^{2}}{\epsilon_{0}S|\vec{Q}|} \nonumber
\end{eqnarray}
The density fluctuation operators for electrons $\hat{\rho }_{e}^{R}(\vec{Q})$ and for holes $\hat{\rho }_{h}^{R}(\vec{Q})$ are determined below:
\begin{eqnarray}
 &\hat{\rho }_{e}^{R}(\vec{Q})=A(Q)\sum\limits_{t}{{{e}^{i{{Q}_{y}}t{{l}^{2}}}}a_{t-\frac{{{Q}_{x}}}{2}}^{\dagger }}{{a}_{t+\frac{{{Q}_{x}}}{2}}}=A(Q)\hat{\rho }_{e}^{{}}(\vec{Q}),& \nonumber\\
& \hat{\rho }_{h}^{R}(\vec{Q})=B(Q)\sum\limits_{t}{{{e}^{i{{Q}_{y}}t{{l}^{2}}}}b_{t+\frac{{{Q}_{x}}}{2}}^{\dagger }}{{b}_{t-\frac{{{Q}_{x}}}{2}}}=B(Q)\hat{\rho }_{h}^{{}}(\vec{Q}), &\nonumber\\
&A(Q)={{\left| {{a}_{0}} \right|}^{2}}+\left( 1-\frac{{{Q}^{2}}{{l}^{2}}}{2} \right){{\left| {{b}_{1}} \right|}^{2}},& \\
&B(Q)={{\left| {{d}_{0}} \right|}^{2}}+\left( 1-\frac{{{Q}^{2}}{{l}^{2}}\left( {{Q}^{2}}{{l}^{2}}-12 \right)\left( {{Q}^{2}}{{l}^{2}}-6 \right)}{48} \right){{\left| {{c}_{3}} \right|}^{2}},& \nonumber\\
&\hat{N}_{e}=\hat{\rho }_{e}^{R}(0),\hat{N}_{h}=\hat{\rho }_{h}^{R}(0).& \nonumber
\end{eqnarray}
They are expressed through the Fermi creation and annihilation operators $a_{p}^{\dagger },{a}_{p}$ for electrons and $b_{p}^{\dagger },{b}_{p}$ for holes. The e-h operators depend on two quantum numbers. Only the electrons and holes on the lowest Landau levels ${{n}_{e}}={{n}_{h}}=0$ are considered and their notations are dropped. The quantum number p denotes the N-fold degeneracy of the Landau levels in the Landau gauge. N equals to $\frac{S}{2\pi {{l}^{2}}}$ where S is the surface layer area and l is the magnetic length $l^{2}=\frac{\hbar c}{eH}$. $a_{0},b_{1},d_{0},c_{3}$ are the coefficients which includes the spin-orbit interaction [17].\newline
The operators (3) obey to the following commutation relations most of which were discussed for the first time in the paper [5, 18]
\begin{eqnarray}
&\left[ \hat{\rho }_{e}^{R}(\vec{Q}),\hat{\rho }_{e}^{R}(\vec{P}) \right]=2i\sin \left( \frac{{{\left[ \vec{P}\times \vec{Q} \right]}_{z}}{{l}^{2}}}{2} \right)\hat{\rho }_{e}^{{}}(\vec{P}+\vec{Q})A(\vec{P})A(\vec{Q}),& \nonumber\\
&\left[ \hat{\rho }_{h}^{R}(\vec{Q}),\hat{\rho }_{h}^{R}(\vec{P}) \right]=-2i\sin \left( \frac{{{\left[ \vec{P}\times \vec{Q} \right]}_{z}}{{l}^{2}}}{2} \right)\hat{\rho }_{h}^{{}}(\vec{P}+\vec{Q})B(\vec{P})B(\vec{Q}),& \nonumber \\
&\left[ \hat{\rho }_{e}^{R}(\vec{Q}),\hat{\rho }_{h}^{R}(\vec{P}) \right]=0 &
\end{eqnarray}
The density fluctuation operators (3) with different wave vectors $\vec{P}$ and $\vec{Q}$ do not commute. These properties considerably influence on the structure of the motion equations for the operators and determine new aspects of the 2D electron-hole (e-h) physics. The motion equations for the operators $\hat{\rho }_{e}^{R}(\vec{Q})$ and $\hat{\rho }_{h}^{R}(\vec{Q})$can be written in the form:
\begin{eqnarray}
&i\hbar \frac{d\hat{\rho }_{e}^{R}(\vec{P})}{dt}=[\hat{\rho }_{e}^{R},\hat{H}]=i\sum\limits_{{\vec{Q}}}{{}}{{W}_{{\vec{Q}}}}\sin \left( \frac{{{[\vec{P}\times \vec{Q}]}_{z}}{{l}^{2}}}{2} \right)\times & \nonumber\\
&\times [\hat{\rho }_{e}^{{}}(\vec{Q})\hat{\rho }_{e}^{{}}(\vec{P}-\vec{Q})+\hat{\rho }_{e}^{{}}(\vec{P}-\vec{Q})\hat{\rho }_{e}^{{}}(\vec{Q})]A(\vec{Q})A(\vec{P})-& \nonumber\\
&-2i\sum\limits_{{\vec{Q}}}{{}}{{W}_{{\vec{Q}}}}\sin \left( \frac{{{[\vec{P}\times \vec{Q}]}_{z}}{{l}^{2}}}{2} \right)\hat{\rho }_{e}^{{}}(\vec{P}-\vec{Q})\hat{\rho }_{h}^{{}}(\vec{Q})A(\vec{P})B(\vec{Q}), & \nonumber \\
&i\hbar \frac{d\hat{\rho }_{h}^{R}(\vec{P})}{dt}=[\hat{\rho }_{h}^{R},\hat{H}]=-i\sum\limits_{{\vec{Q}}}{{}}{{W}_{{\vec{Q}}}}\sin \left( \frac{{{[\vec{P}\times \vec{Q}]}_{z}}{{l}^{2}}}{2} \right)\times & \nonumber\\
&\times [\hat{\rho }_{h}^{{}}(\vec{Q})\hat{\rho }_{h}^{{}}(\vec{P}-\vec{Q})+\hat{\rho }_{h}^{{}}(\vec{P}-\vec{Q})\hat{\rho }_{h}^{{}}(\vec{Q})]B(\vec{Q})B(\vec{P})+ &\\
&+2i\sum\limits_{{\vec{Q}}}{{}}{{W}_{{\vec{Q}}}}\sin \left( \frac{{{[\vec{P}\times \vec{Q}]}_{z}}{{l}^{2}}}{2} \right)\hat{\rho }_{h}^{{}}(\vec{P}-\vec{Q})\hat{\rho }_{e}^{{}}(\vec{Q})B(\vec{P})A(\vec{Q}). & \nonumber
\end{eqnarray}
Following the equations of motion (5) we have introduced two interconnected Green's functions $G_{1j}(\vec{P},t)$  as well as their Fourier transformations $G_{1j}(\vec{P},\omega)$ at $T=0$  [19, 20] of the type $G_{1j}(\vec{P},t)=\langle\langle X_{j}(\vec{P},t);Y(0,t)\rangle\rangle$, where $X_{1}(\vec{P},t)=\hat{\rho }_{e}^{R}(\vec{P}), X_{2}(\vec{P},t)=\hat{\rho }_{h}^{R}(\vec{P})$, and an arbitrary operator $Y(0,t)$ because its choice does not influence on the energy spectrum of the system. These Green's function can be named as one-operator Green's functions. At the right hand sides of the corresponding equations of motion there is a second generation of the two-operator Green's functions. A second generation of equations of motion derived for them containing in their right hand sides the three-operator Green's functions. Following the procedure proposed by Zubarev [20] the truncation of the chains of equations of motion was made and the three-operator Green's functions were presented as a products of one-operator Green's functions $G_{1j}(\vec{P},\omega) $ multiplied by the averages of the type $\left\langle \rho _{e}^{R}(\vec{Q})\rho _{e}^{R}(-\vec{Q}) \right\rangle $ or $\left\langle \rho _{h}^{R}(\vec{Q})\rho _{h}^{R}(-\vec{Q}) \right\rangle $.\newline
The averages $\left\langle \rho _{e}^{R}(\vec{Q})\rho _{e}^{R}(-\vec{Q}) \right\rangle $ and $\left\langle \rho _{h}^{R}(\vec{Q})\rho _{h}^{R}(-\vec{Q}) \right\rangle $ are calculated when the ground state represents the EHL at the temperature $T=0$. This state is characterized by the average occupation number for electrons and holes $\left\langle a_{p}^{\dagger }a_{p}^{{}} \right\rangle =\left\langle b_{p}^{\dagger }b_{p}^{{}} \right\rangle ={{v}^{2}}$ and leads to the values:
\begin{eqnarray}
\left\langle \hat{\rho }_{e}^{R}(\vec{Q})\hat{\rho }_{e}^{R}(-\vec{Q}) \right\rangle =N{{v}^{2}}(1-{{v}^{2}}){{A}^{2}}(\vec{Q}), \nonumber\\
\left\langle \hat{\rho }_{h}^{R}(\vec{Q})\hat{\rho }_{h}^{R}(-\vec{Q}) \right\rangle =N{{v}^{2}}(1-{{v}^{2}}){{B}^{2}}(\vec{Q}), \\
\left\langle \hat{\rho }_{e}^{R}(\vec{Q})\hat{\rho }_{h}^{R}(\vec{Q}) \right\rangle =0.\nonumber
\end{eqnarray}
The Zubarev procedure is equivalent to a perturbation theory with a small parameter of the type $v^{2}(1-v^{2})$
\section{Self-energy parts and the dispersion laws}
The closed system of Dyson equations has the form:
\begin{equation}
\sum\limits_{j=1}^{2}{{}}{{G}_{1j}}(\vec{P},\omega ){{\Sigma }_{ij}}(\vec{P},\omega )={{C}_{1j}},i=1,2
\end{equation}
There are 4 different components of the self-energy parts $\Sigma_{ij}(\vec{P},\omega )$ forming a $2\times 2$ matrix. They are:
\begin{eqnarray}
&{{\sum }_{11}}(\vec{P},\omega )=\hbar \omega +i\delta -\frac{4}{\hbar \omega +i\delta }\sum\limits_{{\vec{Q}}}{{}}{{W}_{{\vec{Q}}}}{{\sin }^{2}}\left( \frac{{{[\vec{P}\times \vec{Q}]}_{z}}{{l}^{2}}}{2} \right)\times  &\nonumber\\
&\times \{{{W}_{{\vec{Q}}}}\left\langle \rho _{e}^{R}(\vec{Q})\rho _{e}^{R}(-\vec{Q}) \right\rangle {{A}^{4}}(\vec{Q})A(\vec{P})+& \nonumber\\
&+{{W}_{{\vec{Q}}}}\left\langle \rho _{h}^{R}(\vec{Q})\rho _{h}^{R}(-\vec{Q}) \right\rangle {{A}^{2}}(\vec{Q})A(\vec{P}){{B}^{2}}(\vec{Q})-& \nonumber\\
&-{{W}_{\vec{P}-\vec{Q}}}\left\langle \rho _{e}^{R}(\vec{P}-\vec{Q})\rho _{e}^{R}(-\vec{Q}+\vec{P}) \right\rangle {{A}^{2}}(\vec{Q})A(\vec{P}){{A}^{2}}(\vec{P}-\vec{Q})\},& \nonumber\\
&{{\sum }_{22}}(\vec{P},\omega )=\hbar \omega +i\delta -\frac{4}{\hbar \omega +i\delta }\sum\limits_{{\vec{Q}}}{{}}{{W}_{{\vec{Q}}}}{{\sin }^{2}}\left( \frac{{{[\vec{P}\times \vec{Q}]}_{z}}{{l}^{2}}}{2} \right)\times  &\nonumber\\
&\times \{{{W}_{{\vec{Q}}}}\left\langle \rho _{h}^{R}(\vec{Q})\rho _{h}^{R}(-\vec{Q}) \right\rangle {{B}^{4}}(\vec{Q})B(\vec{P})+ &\nonumber\\
&+{{W}_{{\vec{Q}}}}\left\langle \rho _{e}^{R}(\vec{Q})\rho _{e}^{R}(-\vec{Q}) \right\rangle {{A}^{2}}(\vec{Q})B(\vec{P}){{B}^{2}}(\vec{Q})- &\nonumber\\
&-{{W}_{\vec{P}-\vec{Q}}}\left\langle \rho _{h}^{R}(\vec{P}-\vec{Q})\rho _{h}^{R}(-\vec{Q}+\vec{P}) \right\rangle {{B}^{2}}(\vec{Q})B(\vec{P}){{B}^{2}}(\vec{P}-\vec{Q})\}, &\nonumber\\
&{{\sum }_{12}}(\vec{P},\omega )=-\frac{4}{\hbar \omega +i\delta }\sum\limits_{{\vec{Q}}}{{}}{{W}_{{\vec{Q}}}}{{W}_{\vec{P}-\vec{Q}}}{{\sin }^{2}}\left( \frac{{{[\vec{P}\times \vec{Q}]}_{z}}{{l}^{2}}}{2} \right)\times  &\\
&\left\langle \rho _{e}^{R}(\vec{P}-\vec{Q})\rho _{e}^{R}(-\vec{Q}+\vec{P}) \right\rangle A(\vec{P})A(\vec{Q})B(\vec{Q})A(\vec{P}-\vec{Q})B(\vec{P}-\vec{Q}), &\nonumber\\
&{{\sum }_{21}}(\vec{P},\omega )=-\frac{4}{\hbar \omega +i\delta }\sum\limits_{{\vec{Q}}}{{}}{{W}_{{\vec{Q}}}}{{W}_{\vec{P}-\vec{Q}}}{{\sin }^{2}}\left( \frac{{{[\vec{P}\times \vec{Q}]}_{z}}{{l}^{2}}}{2} \right)\times & \nonumber\\
&\left\langle \rho _{h}^{R}(\vec{P}-\vec{Q})\rho _{h}^{R}(-\vec{Q}+\vec{P}) \right\rangle B(\vec{P}-\vec{Q})B(\vec{P})B(\vec{Q})A(\vec{Q})A(\vec{P}-\vec{Q}).&\nonumber
\end{eqnarray}
As one can see that all the self-energy parts have only infinitesimal imaginary parts $i\delta $, which means that the elementary excitations are without damping in the given approximation. \newline
The cumbersome dispersion equation for plasma excitations is expressed in general form by the determinant equation:
\begin{equation}
\det \left| {{\Sigma }_{ij}}(\vec{P},\omega ) \right|=0
\end{equation}
and has the form:
\begin{equation}
{{\Sigma }_{11}}(\vec{P},\omega ){{\Sigma }_{22}}(\vec{P},\omega )-{{\Sigma }_{12}}(\vec{P},\omega ){{\Sigma }_{21}}(\vec{P},\omega )=0
\end{equation}
This result well agrees with the result of [12]. Indeed, if we assume that there is no spin-orbit interaction then by [17] ${{\left| {{a}_{0}} \right|}^{2}}={{\left| {{d}_{0}} \right|}^{2}}=1$ and ${{\left| {{c}_{3}} \right|}^{2}}={{\left| {{b}_{1}} \right|}^{2}}=0$ , and we will get exactly the same expression for the optical and acoustical plasmon oscillations [12].
Solving the equation (10) can be deduced the energy spectrums characterized by the dispersion laws:
\begin{eqnarray}
&{{(\hbar {{\omega }_{1}}(Pl))}^{2}}=4\sum\limits_{{\vec{Q}}}{{}}W_{{\vec{Q}}}^{2}{{\sin }^{2}}\left( \frac{{{[\vec{P}\times \vec{Q}]}_{z}}{{l}^{2}}}{2} \right)\times & \nonumber\\
&\times [\left\langle \rho _{e}^{R}(\vec{Q})\rho _{e}^{R}(-\vec{Q}) \right\rangle {{A}^{4}}(\vec{Q})A(\vec{P})+& \nonumber\\
&+\left\langle \rho _{h}^{R}(\vec{Q})\rho _{h}^{R}(-\vec{Q}) \right\rangle {{A}^{2}}(\vec{Q})A(\vec{P}){{B}^{2}}(\vec{Q})]- &\nonumber\\
&-4\sum\limits_{{\vec{Q}}}{{}}W_{{\vec{Q}}}^{{}}W_{\vec{P}-\vec{Q}}^{{}}{{\sin }^{2}}\left( \frac{{{[\vec{P}\times \vec{Q}]}_{z}}{{l}^{2}}}{2} \right)\times  &\nonumber\\
&\times \left\langle \rho _{e}^{R}(\vec{P}-\vec{Q})\rho _{e}^{R}(-\vec{P}+\vec{Q}) \right\rangle {{A}^{2}}(\vec{Q}){{A}^{2}}(\vec{P}-\vec{Q})A(\vec{P}), &\nonumber\\
&{{(\hbar {{\omega }_{2}}(Pl))}^{2}}=4\sum\limits_{{\vec{Q}}}{{}}W_{{\vec{Q}}}^{2}{{\sin }^{2}}\left( \frac{{{[\vec{P}\times \vec{Q}]}_{z}}{{l}^{2}}}{2} \right)\times&  \\
&\times [\left\langle \rho _{h}^{R}(\vec{Q})\rho _{h}^{R}(-\vec{Q}) \right\rangle {{B}^{4}}(\vec{Q})B(\vec{P})+& \nonumber\\
&+\left\langle \rho _{e}^{R}(\vec{Q})\rho _{e}^{R}(-\vec{Q}) \right\rangle {{A}^{2}}(\vec{Q})B(\vec{P}){{B}^{2}}(\vec{Q})]- &\nonumber\\
&-4\sum\limits_{{\vec{Q}}}{{}}W_{{\vec{Q}}}^{{}}W_{\vec{P}-\vec{Q}}^{{}}{{\sin }^{2}}\left( \frac{{{[\vec{P}\times \vec{Q}]}_{z}}{{l}^{2}}}{2} \right)\times  &\nonumber\\
&\times \left\langle \rho _{h}^{R}(\vec{P}-\vec{Q})\rho _{h}^{R}(-\vec{P}+\vec{Q}) \right\rangle {{B}^{2}}(\vec{Q}){{B}^{2}}(\vec{P}-\vec{Q})B(\vec{P}). &\nonumber
\end{eqnarray}
This is dispersion relations for acoustical plasmon $\hbar {{\omega }_{1}}(Pl)$ and for the optical plasmon $\hbar {{\omega }_{2}}(Pl)$. The dispersion law for the acoustical plasmon has a linear dependence on the wave vector in the range of long wavelength and tends to constant value at great values of Pl. The dispersion relation for optical plasmons has a quadratic dependence at small values of Pl and a similar behavior at great values of Pl as in the case of acoustical plasmons. The difference between them is essential only in the range of intermediary values of Pl. The dispersion laws for acoustical and optical plasmon type excitations are represented in the Fig. 1 for the GaAs-type crystal being under the influence of high magnetic and electric fields: ${{E}_{z}}=10kV/cm$ and $H=10T$. Filling factor ${{v}^{2}}$ are equal to 1/3.
\section{Conclusions}
The plasma oscillations of the 2D electron-hole liquid formed on the surface of the layer subjected to the action of a strong perpendicular magnetic field  and under influence Rashba spin-orbit coupling are characterized by two branches of the energy spectrum. One of them is the acoustical plasmon type branch with the linear dispersion law in the range of small wave vectors and monotonically increasing with saturation behavior at higher wave vectors. The second branch of the elementary excitations is an optical-plasmon branch with quadratic dispersion law at small wave vectors and with a similar behavior as the acoustical branch at higher wave vectors. Results obtained in the paper are in good agreement with previous work. In the limit, excluding the Rashba spin-orbit coupling we get exactly the results obtained in [12].
\begin{figure}
\resizebox{0.48\textwidth}{!}{%
  \includegraphics{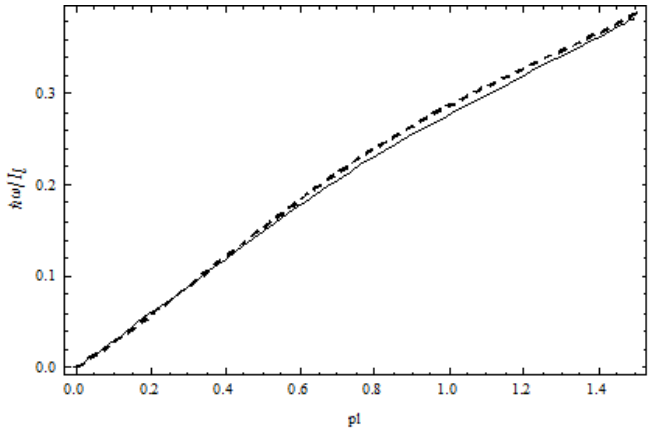}
}
\caption{The dispersion law for the acoustical(solid line) and optical(dotted line) plasmon branches. Filling factor ${{v}^{2}}$ is equal to 1/3. Magnetic and electric fields are equals to ${{E}_{z}}=10kV/cm$ and $H=10T$.}
\label{fig:1}       
\end{figure}

\end{document}